\documentclass[11pt,nofootinbib,showpacs]{revtex4}
\usepackage{amsfonts,amssymb,amsmath,graphicx}
\def\[{\left[}
\def\]{\right]}
\def\({\left(}
\def\){\right)}

\begin{document}

\title{On Hamiltonian intermittency in equal mass three-body problem}

\author{S.A. Pavluchenko}
\affiliation{Special Astrophysical Observatory, Russian Academy of
Sciences, Nizhnij Arkhyz, 369167 Russia}

\begin{abstract}
We demonstrate that both kinds of the Hamiltonian intermittency
exert an influence on the disruption statistics in the equal mass
three-body problem. Studying initially-resting triple systems we
found a narrow region in the vicinity of the strong chaos, where
the influence of the second kind Hamiltonian intermittency
($T_d^{-3/2}$) trajectories cause the integral distribution to
distort enough to be detected. We fitted the integral distribution
with both power-laws ($T_d^{-3/2}$ and $T_d^{-2/3}$) taken into
account, and found an excellent agreement between the fit and
observed integral distribution.
\end{abstract}


\maketitle
\twocolumngrid

\section{Introduction}

The three-body problem, although being the simplest of the general
$N$-body problem, does not have an analytical solution in the
general case, which makes it possible to gain an understanding of
its dynamics mostly through the means of numerical studies. Those
studies started even before any computers were developed and trace
back to the beginning of the 20th century~\cite{1st3body}. With
the development of computers the effectiveness of these studies
increased, but, due to the insufficient amount of data, this does
not led to an immediate breakthrough in the understanding of the
disruption process. Valtonen~\cite{valtonen} assumed that the tail
of the disruption distribution should be exponential, though this
assumption conflicts with the earlier theoretical result by
\mbox{Agekian et al.~\cite{Agekian1983}} that the average
life-time of an isolated triple system should be infinite. Later
Mikkola and Tanikawa~\cite{MT07} found an exponential tail in the
disruption statistics of the equal mass three-body problem.

Apart from this, Shevchenko~\cite{sh10} has recently demonstrated
that in the hierarchical three-body problem the decay of the
survival probability is heavy-tailed with a power index equal to
--2/3, which corresponds to the first kind of Hamiltonian
intermittency; the power-law tails in the equal mass three-body
problem with indices close to that value were recently reported by
Orlov et al.~\cite{Orlovetal2010} (they also appear in the decay
of the survival probability of the exited atoms~\cite{atoms}). The
second kind of the Hamiltonian intermittency predicts the
existence of a power-law tails with index equal to --3/2; a power
index close to this value was reported by Shevchenko and
Scholl~\cite{shsch97} in the 3/1 Jovian resonance
(Sun--Jupiter--Asteroid problem). Other than that, it appears that
power-law tails are common for the disruption statistics of the
Hamiltonian systems of different nature (see~\cite{powerlaws} and
references therein).

As it was mentioned, both kinds of Hamiltonian intermittency were
seen in the three-body problem, hence, they both are ``native'' to
this problem, yet, no one reported seing both of them in one set.
That is why we decided to thoroughly investigate the equal mass
three-body problem and demonstrate that both of them are present.
In the next section we describe our numerical method and the
initial data set, then we report the results and discuss them in
the final section.


\section{Numerical method}

\begin{figure}
\includegraphics[width=0.5\textwidth]{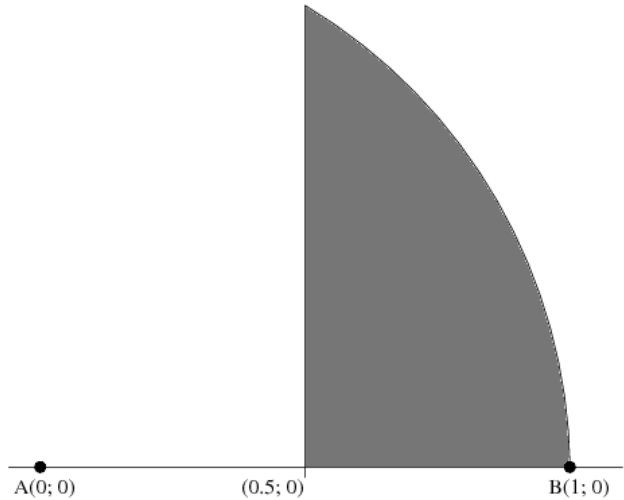}
\caption{Initial positions of the masses: first and second are
located in points A and B while third mass could be placed in any
point in the gray region}\label{fig1}
\end{figure}

To perform our calculations we use the version of the Aarseth-Zare {\it triple} code~\cite{AZ}. 
In contrast with our previous paper~\cite{we11_1} we used a
``standard'' way to define the initial conditions for the
initially-resting triple system, used in \mbox{Agekian et
al.~\cite{Agekian1983}}, where the authors demonstrated that the
approach used cover all possible configurations of three bodies.
It is demonstrated in Fig.\,\ref{fig1} -- that two bodies are
placed in points A and B, while the third body could be placed in
any point in the gray region. The gray region is bounded by the
unit circle centered in A from the right, $x=0.5$ from the left
and $x$-axis from the bottom. Scanning over all possible initial
positions of the third body gives us a complete set of data (as we
use the equal-mass
problem, that is enough; should we use different masses, we have
to permute the masses between A, B, and the ``running'' position,
and use all the resulting time maps together). Initially, all
three masses are at rest, but immediately after they start moving
in the gravitational potential they govern. The calculation of
their motion lasts until the system disrupts (at time $T_d$) or
$T=5000$ is reached; the value for the maximal $T$ is adopted
from~\cite{we11_1}, where we saw that by this time $T_d^{-2/3}$ is
clearly reached in the equal-mass case, our current research
proved that as well. The disruption of the system is determined by
the hyperbolicity (positivity of mechanical energy of the
disrupting binary) at a distance 50 times larger than the current
semi-major axis of the final binary. This way of defining is by
default used in the {\it triple} code that we are using and it is
quite wide-spread in studying the three-body problem in astronomy.

\section{Results}

\begin{figure*}
\includegraphics[width=0.6\textwidth, angle=-90, bb=45 20 345 435, clip]{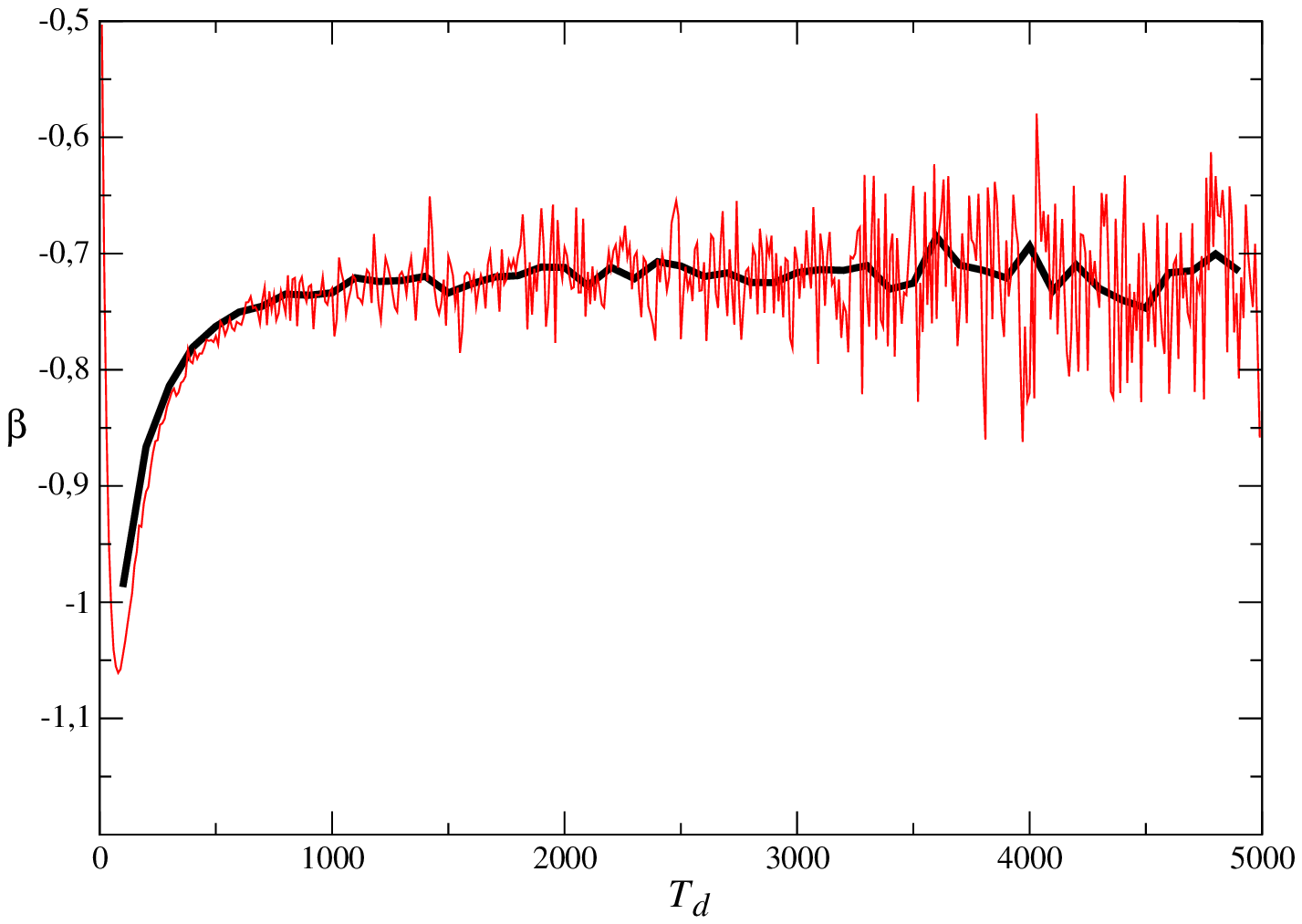}\\
\includegraphics[width=0.6\textwidth, angle=-90, bb=45 20 345 435, clip]{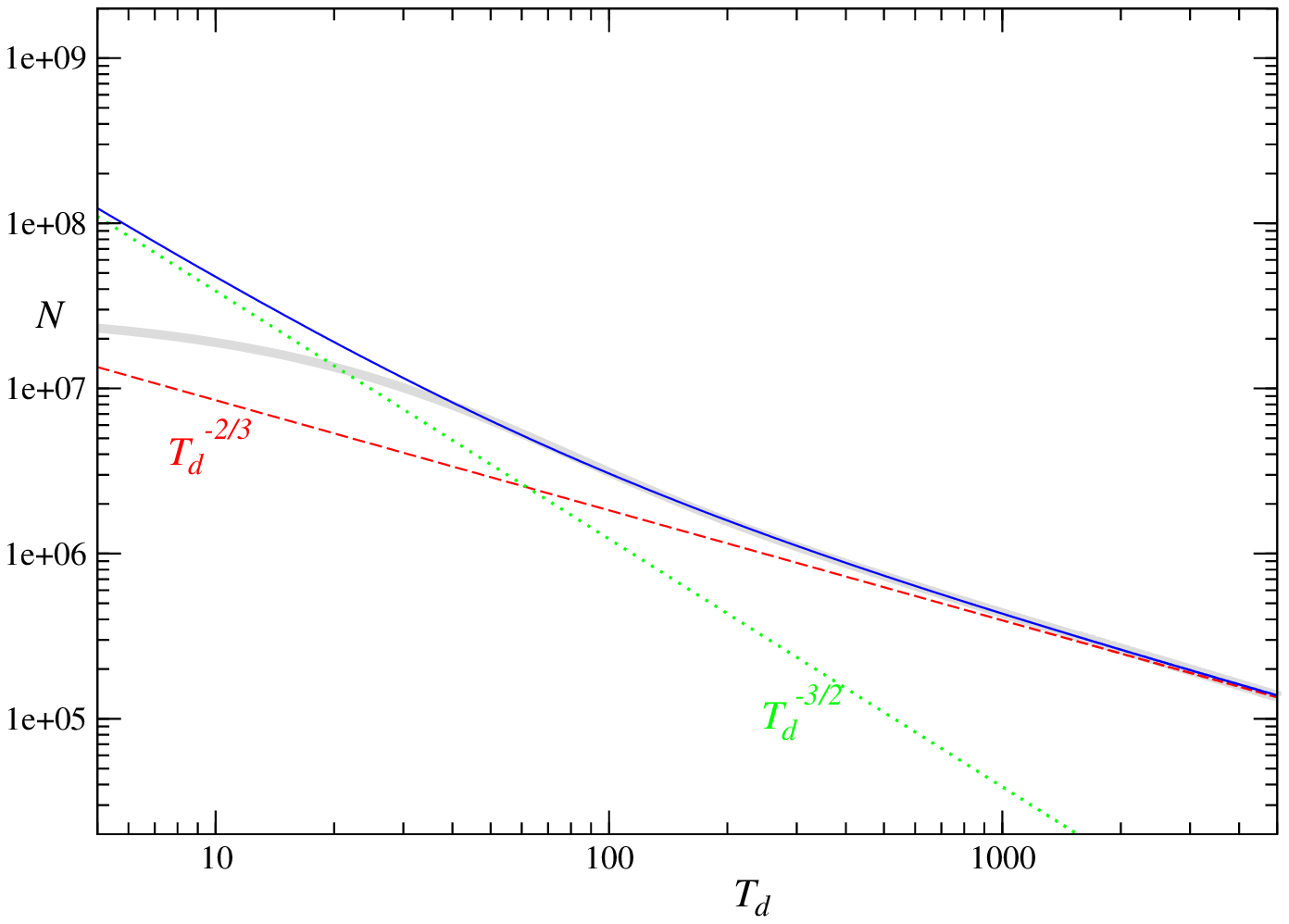}
\caption{Upper panel: the power index of the $F_a(T_d)$ power-law
fit ($F_a(T_d) \propto T_d^{\beta}$). The red (dark gray if
grayscaled) curve corresponds to $\Delta T = 10$ step while the
thick black curve -- to $\Delta T = 100$. Bottom panel: best-fit
of the $F_a(T_d)$ curve with both contributions: the thick gray
line -- $F_a(T_d)$, the green dotted line -- $T_d^{-3/2}$
contribution, the red dashed line -- $T_d^{-2/3}$, the thin blue
line -- the sum of two contributions. See the text for exact
values of powers.} \label{Fa}
\end{figure*}

We have modelled over 30 million initial positions for the third
mass that are evenly-distributed over the gray area in
Fig.\,\ref{fig1}. The resulting $F_a(T_d)$ integral distribution
(the number of trajectories with the disruption time that is
greater than $T_d$) was fitted by the power-law relation $F_a(T_d)
= A \times T_d^{\beta}$ with steps $\Delta T = 10$ and $\Delta T =
100$; the result is presented in the upper panel of
Fig.\,\ref{Fa}: $\Delta T = 10$ as the red (dark gray if
grayscaled) and $\Delta T = 100$ as the thick black line. The
strong noise of the first curve is caused by an insufficient
number of data points in each bin ($\Delta T = 10$), one can
easily see that the second curve is much smoother and still, it
has variations. They both approach $\beta = -0.7 \ldots -0.75$,
which is pretty close to $\beta = -2/3$ that is expected from the
first type of the Hamiltonian intermittency. On the other hand,
the minimum at low $T_d$ points to $\beta = -3/2$ (although, it
does not reach it, ending at $\beta \approx -1.06$) that is
expected from the second type of the Hamiltonian intermittency,
hence we investigate it further on. One note needs to be taken --
with $\Delta T \gtrsim 200\ldots 400$ minimum at low $T_d$ could
be easily missed and we believe that large $\Delta T$ was a reason
why $\beta = -3/2$ was not detected in this problem earlier.



Then we fit the $F_a(T_d)$ curve as a sum of both contributions
$F_a(T_d) = A_1 \times T_d^{\beta_1} + A_2 \times T_d^{\beta_2}$
and this fit gives us an excellent result: \mbox{$A_1 =
3.93812\times 10^7 \pm 5.297\times 10^5 ~(1.345\%)$}, \mbox{$A_2 =
1.2327\times 10^9 \pm 3.707\times 10^7 ~(3.007\%)$},
\mbox{$\beta_1 = -0.666399 \pm 0.001526 ~(0.229\%)$}, and
\mbox{$\beta_2 = -1.501 \pm 0.008454 ~(0.5633\%)$}. One can see
that the powers are almost exactly equal to the predicted values;
in the bottom panel of Fig.\,\ref{Fa} we presented the $F_a(T_d)$
curve (thick gray), $T_d^{-3/2}$ contribution (dotted green),
$T_d^{-2/3}$ (red dashed), and the sum of two contributions (thin
blue). One can see that the resulting curve fits $F_a(T_d)$ quite
well, while $T_d^{-2/3}$ alone coincide with $F_a(T_d)$ only
starting from thousands of $T_d$.

%

\section{Conclusions}

We have considered the three-body problem with unit masses that
are initially at rest and demonstrated that both terms --
$T_d^{-2/3}$ and $T_d^{-3/2}$ -- contribute into the statistics of
the disruption. The reason why $T_d^{-3/2}$ contribution remained
undetected in the equal-mass problem is because the corresponding
trajectories are stuck to the strong chaos region at low $T_d$.
The chaotic pattern ends at the low values of $T_d$, so the
trajectories that are stuck to its border manifest themselves
strongest at low $T_d$ as well (though, not solely, since it is
owing to the influence of $T_d^{-3/2}$ contribution that the power
of $T_d^{-2/3}$ is not found exactly). Fitting the $F_a(T_d)$
curve clearly demonstrated the presence of both the contributions
($T_d^{-2/3}$ and $T_d^{-3/2}$), and the powers were confirmed
with great precision. We believe that the reason why the power
indices found in the earlier papers on this subject
(e.g.,\cite{Orlovetal2010}) were somehow lower than the expected
$-2/3$ was the interference from the $T_d^{-3/2}$ term. Also, high
values for the maximal integration time usually lead to high
values of $\Delta T$ -- the size of bin to plot the distribution.
If $\Delta T \gtrsim 200 \ldots 400$, then the whole deep in upper
panel of Fig.\,\ref{Fa} would be inside one bin, and so it could
be undetected -- we believe that this was, at least, partially, a
reason why both indices simultaneously remained undetected.

\section{Acknowledgments}

The author is grateful to Alexey Toporensky and Ivan Shevchenko
for fruitful discussions and to Sergey Karpov and Margarita
Khabibullina for the computational help.

\end{document}